# Ranking power spectra: A proof of concept

Zhenning Mei[1], Xilin Yu[1], Chen Chen[1] and Wei Chen[1]

[1] Center for Intelligent Medical Electronics (CIME), Fudan University, Shanghai China

**Abstract**

*Objective*: To characterize the irregularity of the spectrum of a signal, spectral entropy is a widely adopted measure. However, such a metric is invariant under any permutation of the estimations of the powers of individual frequency components on a predefined grid. This erases the order structure inherent in the spectrum which is also an important aspect of irregularity of the signal. To disentangle the order structure and extract meaningful information from raw digital signal, novel analysis method is necessary.
*Approach*: A novel method to depict the order structure by simply ranking power estimations on frequency grid of a evenly spaced signal is proposed. Two descriptors mapping real- and vector-valued power spectrum estimation of a signal into scalar value are defined in a heuristic manner. By definition, the proposed descriptor is capable of distinguishing signals with identical spectrum entropies.
*Main Results*: The proposed descriptor showed its potential in diverse problems. Significant ($p<0.001$) differences were observed from brain signals and surface electromyography of different pathological/physiological states. Drastic change accompanied by the alteration of the underlying process of signals enables it as candidate feature for seizure detection and endpoint detection in speech signal.
*Significance*: This letter explores the previously ignored order structure in the spectrum of physiological signal. We take one step forward along this direction by proposing two computationally efficient descriptors with guaranteed information gain. As far as the authors are concerned, this is the first work revealing the effectiveness of the order structure in the spectrum in physiological signal processing.

Keywords: biomedical signal processing, order structure, spectral entropy

## 1. Introduction

Many real-world signals including physiological signals are irregular on some aspects. They are neither purely periodic nor can they be expressed by an analytic formula. The inherent irregularities imply the uncertainty during the evolution of the underlying process from which the signals are observed. The uncertainty enables information transfer but also limits the predictability of those signals. The unpredictability of signal in time domain makes researchers have to toil in frequency domain. Fourier transform (FT) bridges signals in original space (time domain) with their representations in dual space (frequency domain) by decomposing a signal satisfying some weak constraints into infinitely many periodic components, which has numerous applications in signal processing.

For most of the real-world applications, finite samples drawn (usually evenly spaced) from a continuous random process cannot give us full information about the process' evolution but only a discrete depiction. FT was adapted into discrete fourier transform (DFT) for such scenarios (Cooley et al., 1969). And line spectrum wherein the total energy of the signal distributes on only few frequency components is rarely encountered among physiological signals due to the inherent irregularities therein.

To characterize the irregularity of digital signals in frequency domain. Spectral entropy is introduced analogous to the Shannon entropy in information theory (Shannon, 1948). The estimations on the frequency grid are firstly divided by the total power and then a list of proxies in the form of probabilities whose sum is 1 is obtained. Then the Shannon entropy formula which is the negative sum of probability-weighted log probabilities map those proxies into a quantity representing the irregularity of energy distribution on frequency domain. Under this perspective, a flat spectrum has maximal spectral entropy and the spectrum of a single-frequency signal has minimal spectral entropy, which is zero. Spectral entropy has been applied in diverse areas, including endpoint detection in speech segmentation (Bing-Fei Wu and Kun-Ching Wang, 2005), spectrum sensing in cognitive radio (Zhang et al., 2010). And it has also served as the base of a famous inductive bias, maximum entropy (Smith and Grandy, 1985), which is widely adopted for spectrum estimation of some kinds of physiological signals like electroencephalogram (EEG).

Although spectral entropy is well-defined and can be computed efficiently by Fast Fourier Transform (FFT), it is difficult to relate spectral entropy with other interpretable properties of interest of original signal especially when take no account of the overwhelming endorsement from its counterpart being the foundational concept in information theory which quantifies the *uncertainty*. Furthermore, it is apparent that the spectral entropy ignored the order information since the power estimations are arranged on the frequency grid having intrinsic partial order structure. Any permutations of these values on the grid yields a same spectral entropy. But obviously, the representations of those signals in time domain can look very different.

To incorporate the order information carried by the power spectrum is guided by the following belief. The normal operations of any system (biological/electromechanical, etc.) are impossible without the processing of information through some physical/chemical process. It could be the signaling between different modules within the system or the communications between the system as a whole and the external environment. Information transfers happened in those scenarios are accomplished with the help of carrier signals of particular forms with nontrivial structures beared on their spectra. And only limited frequency precision of the control and recognition of those signals are practical for real systems. So it is unreasonable for natural systems that have gone through long-term evolution or well-designed artificial systems to arrange the internal signals responsible for different functions close with each other in frequency domain within certain time window. Otherwise the efficient transfer of information could be degraded. And frequency divided multiplex (Zhang et al., 2010) in modern communication systems can be considered as a living example of this belief.

Therefore, if we use power estimations on the frequency grid as proxies of the *intensities* of activities corresponding to those frequencies, it seems reasonable to infer that the energy distributed on neighboring rather than remote frequency grids are more likely caused by same function. The alpha band activitites (8-13 Hz) which can be interrupted by visual perception tasks in human's EEG is one of the examples.

To sum up, if we want to develop a metric characterizing aforementioned structural irregularities of the power spectra. It was supposed to assign a larger value to a signal wherein the frequency components with *similar* intensities are distributed far apart from rather than close to each other. And the *similarities* of intensities can be reflected (partially) by the relative order of power estimates on discrete frequency grid. That is why the order information in the spectrum can shed new light on the structure aspects of signal and how the order information is incorporated into analysis.

In this letter, we explore the effectiveness of the order information carried by the power spectra of signals. Given the motivation illustrated above, in part 2 we provide details about the way we used to evaluate it. In part 3 we present several use cases to justify the effectiveness of our preliminary approach and more importantly, the promising potential to find some new research niche in the field of physiological signal processing. Finally, discussion about the limitations of our work and future directions are followed in part 4.

## 2. Method

Given an equally spaced, real-valued digital signal $s$, we assume the length of $s$ is an even number $2N$, for simplicity. Then DFT is applied to $s$ and a complex-valued vector $\hat{s}$ of dimension $2N$ is obtained as follows :

$$\hat{s}_k = \sum_{i=0}^{2N-1} s_i \cdot e^{-j\cdot\frac{i\cdot k\cdot 2\pi}{2N}}, \quad k = 0,1,\dots,2N-1 \quad (1)$$

Due to the conjugate symmetry of $\hat{s}$, we take the square of the modulus of the first half of $\hat{s}_k$ and get $P_s \in R^N$. Thanks to the *Parseval* identity, the 1-norm of $P_s$ equals to the energy of $s$ up to factor 1/2. Although $P_s$ has a dimension of energy instead of power, the constant factor having a dimension of time does not change the relative ordinal relations between its components. So we just use $P_s$ as the estimations of power on normalized frequency range $[0, \pi]$, whereby the $k_{th}$ component of $P_s$ is the estimation of signal's power on grid point $(k-1)\cdot \pi/N$.

Now let assume again that every two components of $P_s$ are different from each other so we can rank these $N$ components without any ambiguity in ascending/descending order.

These grid points have an intrinsic partial order structure from low frequency range to high frequency range. So we get eigen-triple for these grid points:

$$\begin{pmatrix} 1 & 2 & 3 & \cdots & N \\ P_s(1) & P_s(2) & P_s(3) & \cdots & P_s(N) \\ \sigma(1) & \sigma(2) & \sigma(3) & \cdots & \sigma(N) \end{pmatrix} \quad (2)$$

The first row indicates the grid points by their location on frequency range. The second row contains the corresponding power estimations. The third row contains the relative order of corresponding power estimation among all estimations, denoted by $\sigma(\cdot)$. Since no duplicated values in $P_s$ are assumed, $\{\sigma(i)\}_{i=1}^{N}$ will traverse number set $\{1,2,3,\ldots,N\}$.

The first two rows of (2) are just a kind of representation of traditional power spectrum. Novelty lies in taking the order information, carried in the third row, into consideration.

It should be noted that the first and the third row together have defined a permutation over the natural number set $\{1,2,3,\ldots,N\}$, with its complete detail determined by $s$ implicitly. Remind that spectral entropy is defined in a permutation-invariant way. Such an invariance must be broken down so as to disentangle the order information. Therefore, this permutation *per se* returns the long-overdue ladder to understand structural irregularities of signals under a new perspective.

The sketch of our method is illustrated in Fig. 1. Using the measurements in time domain (Fig. 1(a)), the power estimations on normalized frequency grid with resolution determined by half the length of original signal are obtained (Fig. 1(b)). By ranking these estimations in descending order, we arrange $\sigma_{(i)}^{-1}$ against $i$. As showed in Fig. 1(c), the first stem indicates the location on the frequency grid of the largest power component, and so on. From (b) to (c), we are actually performing a nonlinear stretching while the order information of the spectrum is preserved and calibrated. Then a distance matrix $M$ (in Fig. 1(d)) is induced for every point pair. Here in (c) we define $M_{ij} = M_{ji} = |\sigma_{(i)}^{-1} - \sigma_{(j)}^{-1}|$.

So $M$ is real-symmetric with trace identically equals to 0. The structural aspects of $M$ are reflected in its eigenvalues (Fig. 1(e)). Due to the sophisticated relationships between its entries, its unwise to reshape such a high dimensional object with far lower degrees of freedom into a long vector for pattern recognition. In addition to the eigenvalues, a descriptor named as *Circular Difference Descriptor* ($C_iD$), accounting for the total variation of the locations on frequency grids of frequency components having adjacent intensities is defined as follows, in a heuristic manner:

$$C_iD_N = \frac{1}{N} \cdot \left( |\sigma_{(N)}^{-1} - \sigma_{(1)}^{-1}| + \sum_{i=1}^{N-1} |\sigma_{(i)}^{-1} - \sigma_{(i+1)}^{-1}| \right) \quad (3)$$

The first term makes *Circular Difference* vertible and endows $C_iD$ translational invariance instead of permutational invariance.

Another heuristic descriptor is defined slightly different from $C_iD_N$, named as *Correspondence Difference Descriptor* ($C_oD$). It equals to the 1-norm of the difference of $\sigma_{(i)}^{-1}$ and $i$, aiming to characterize the difference between $\{\sigma^{-1}(i)\}_{i=1}^{N}$ and the perfectly ordered case where $\sigma(i) = i$:

$$C_oD_N = \frac{1}{N} \cdot \sum_{i=1}^{N} |\sigma_{(i)}^{-1} - i| \quad (4)$$

Results from the Monte-Carlo simulation (showed in Supplementary Information) implies that the empirical distributions of $C_iD_N$ and $C_oD_N$ among all $N!$ permutations could well be Gaussian. Although theoretical distributions of $C_iD_N$ and $C_oD_N$ must have bounded supports for finite $N$, but they fit a bell-shaped curve very well, which in theory has unbounded support.

Since permutational invariance is broken herein, $C_iD$ and $C_oD$ actually encode the signal in different ways but both with guaranteed information gain with respect to spectral entropy. To be specific, given $\{\sigma(i)\}_{i=1}^{N}$ without $\{P_s(i)\}_{i=1}^{N}$, the $C_iD$ and $C_oD$ are fixed but the distribution of $\{P_s(i)\}_{i=1}^{N}$ can form widely differed spectra, from almost flat spectrum (e.g. $P_s(i) = k + \sigma_{(i)}^{-1} \cdot \varepsilon$, where $\varepsilon \downarrow 0$ and $k \uparrow \frac{1}{N}$) to almost line spectrum (e.g. $P_s(\sigma_{(1)}^{-1}) = 1 - \frac{N(N-1)}{2} \cdot \varepsilon$ and $P_s(\sigma_{(i)}^{-1}) = \varepsilon \cdot i, i \neq 1, \varepsilon \downarrow 0$). And the corresponding spectral entropy values vary from infinitesimal to maximum possible. On the contrary, given $\{P_s(i)\}_{i=1}^{N}$, any permutation on it yields the same spectral entropy, as mentioned before, but the $C_iD$ and $C_oD$ will absolutely transverse all possible values.

If no *a priori* about the signals' spectra is available, then the equiprobable distribution of $\{\sigma(i)\}_{i=1}^{N}$ is substantially and implicitly pre-assumed. Then under such circumstance, the so-called *Kullback-Liebler Divergence* (KLD) between the proposed descriptors and spectral entropy as different coding schemes are always positive (Kullback and Leibler, 1951), no matter the direction (KLD is lack of symmetry). Such a property is welcomed since it guarantees the nonnegative information gain when using both spectral entropy and proposed descriptor instead of only one of them.

As for the distance matrix with its entries $M_{ij}$ representing the *distance* or *similarity* between point $i$ and $j$, distance measures other than the absolute value can be applied on $\sigma_{(i)}^{-1}$ and $\sigma_{(j)}^{-1}$ to form different distance matrices. Given any distance measure, a topology is induced on this finite set $\{P_s(i)\}_{i=1}^{N}$, based on the coarse-grained, discrete-valued rankings among them. And certainly, more order information is unrevealed yet. For example, $C_iD$ is just the circular difference of the first subdiagonal line of $M$, captures only a portion of full information.

To sum up, by ranking power estimations of signal on a discrete frequency grid, an interesting picture of order structure carried by signal's spectrum is obtained.

## 3. Results

In this section we provide several use cases to show the effectiveness of order information carried by signal's power spectrum in physiological signal processing.

### 3.1 Surface Electromyography (sEMG)

It was found the proposed descriptor may be able to distinguish sEMG signals collected under different actions. A public available dataset containing sEMG recordings from 3 females and 2 males acting 6 different actions, is involved in analysis (Sapsanis et al., 2013). Wilcoxon rank sum test is used to compare the medians of each class.

A representative example is given in Fig. 2 with statistical significance (p<0.001) achieved between the medians among most of comparisons. As for full comparison, Supplementary Information contains all comparisons for remaining subjects.

### 3.2 Electroencephalogram (EEG)

It was also found that the proposed descriptors may be able to distinguish brain signals under different pathological states. A public available dataset, *Bonn Seizure Dataset*, which is widely used as materials for brain signal related pattern recognition and machine learning tasks is employed (Andrzejak et al., 2001). In this dataset, 5 subsets contain 100 recordings each with identical length, sampling frequency and other conditions, collected under different pathological states. Rank sum test used in *3.1* is performed.

Significant (p<0.001) differences between the medians of the values of proposed descriptors corresponding to these 5 subsets were observed in most of the cases, with boxplots given in Fig. 3.

### 3.3 Speech signal

When $N$ is fixed and performing the operator defined in (3) and (4) on moving window mounted on a long signal, we are able to unfold the structural irregularities of signal in a finer time resolution. And thus change point detection is possible.

In Fig. 4 we provide an example of endpoint detection in human speech signals. It can be seen that the start points and stop points of syllables are accompanied by the steep increase/decrese of descriptors' values. In this example, we also found the descriptors defined in (3) and (4) which are purely based on order information, as opposed to spectral entropy which has nothing to do with order information, could become noise vulnerable in some problems. This is due to the amplitudes of $\{P_s(i)\}_{i=1}^N$ are barely removed after transforming into $\{\sigma(i)\}_{i=1}^N$. Consider such a case where a large portion of $\{P_s(i)\}_{i=1}^N$ only accounts for a negligible portion of total energy, then their relative order can vary drastically because of possible noise, so were the descriptors' values. But it seems unreasonable to deem the structure of signal must have changed accordingly.

Therefore, a simple thresholding segmentation trick of total variance, similar to what is usually adopted in principal component analysis is used in this case. The descriptors are calculated based on the first $L$ components $\{\sigma_{(i)}^{-1}\}_{i=1}^L$ whereby $L$ is defined as follows:

$$L = \min\left\{l \mid \sum_{i=1}^{l} P_s(\sigma_{(i)}^{-1}) \geq q \cdot \left(\sum_{i=1}^{N} P_s(i)\right)\right\} \quad (5)$$

The $q \in (0,1]$ is a tunable parameter selecting $L$ largest frequency components accounting for just above a preset portion of total energy. This trick improves the robustness against wide-band weak noise but removes some welcomed properties either. Possible modifications of naive descriptors proposed here will be discussed later.

### 3.4 Amplitude-integrated EEG (aEEG)

Another example validates the effectiveness of proposed method in revealing the temporal evolution of physiological process is founded in the analysis of aEEG (Mastrangelo et al., 2013). aEEG is a kind of condensed sketch of long-term EEG recording. It was believed to be able to reflect long-term trends of brain activities in a horizon suitable for visual inspection and evaluation. It has been widely used for seizure detection in neonates, brain disorder evaluation, etc.

In Fig.5 a segment of EEG drawn from CHB-MIT dataset (Shoeb and Guttag, 2010) is transformed into aEEG first and then similar analysis used in 3.3 is adopted. Ictal period is indicated by colored bar.

## 4. Discussion

Order structure of signal's spectrum is revealed by simplying ranking the power estimations. Several use cases justify that taking that order structure into consideration could contribute valuable information to the processing of physiological signals. The possible applications include serving as candidate features for pattern recognition among signals, change point detection in process tracking for anomaly detection and many more.

The permutation of length $N$ defined by rankings of power estimations on frequency grid has huge capacity ($N!$). Although in practice it is not necessarily that these $N!$ ordinal patterns are equiprobable, the proved information gain under such an assumption is still hopeful to be found in practice. An established metric *permutation entropy* is based on ranking consecutive measurements in time domain and do statistics among a sufficient number of segments (Bandt and Pompe, 2002). The length of such segments must be small otherwise the density estimation will be impractical for time series of reasonable length. Our method delves into the order structure of signal's representation in dual space (frequency domain) instead of original space (time domain). Every point in the dual space is bridged to all points in the time domain through FT, so no one-to-one correspondence exists between original measurements and mapped points in the proposed method, This is also an important distinction.

The proposed descriptors in their original forms could be noise vulnerable but they can be modified using techniques include but not limit what is used here. In practice we observed high correlation between $C_iD$ and $C_oD$, and one could outperform another at times. In addition, the pairwise distances in the distance matrix in Fig.1(d) can be induced in a way other than that used here. Anyway, more fruitful and distinguishable features can be extracted along this way from such a representation with large capacity.

As for future research, we have several proposals.

The first is to establish relationships between the order information given by a recorded digital signal of length $2N$ and that of its sub-signals, obtained by (nonuniformly) downsampling these $2N$ points. Uniformly downsampling is equivalent to fold the power spectrum. Situations will be more sophisticated under nonuniform cases but usually a more flat spectrum with lower frequency resolution is produced. The original signal with its sub-signals together could provide an informative and hierarchical object of study.

The second is to develop distance measures other than the absolute difference of ranks used here. By incorporating both the discrete-valued ranks and the continous-valued power estimations, parameters more robust to broad band noise could be anticipated. Furthermore, could 'ranking' of power spectrum of a *continuous* function (signal) possible in some sense in theory?

The third is about the topology induced from the distance matrix. The distance matrix in Fig.1(d) or the distance defined by possible modified measures, as mentioned above, whereby block structures frequently occur, provides full neighbourhood information of $N$ points on frequency grid. Given such information, we can calculate so-called *persistent homology*, a dominant methodology usually referred as synonym of topological data analysis (TDA), of these $N$ points by computing a series of simplicial complexes with their topological invariants (Rote and Vegter, 2006) and get topological description of signal's power spectrum. That means the order information in spectrum enables a non-trivial embedding method of data points with temporal structure. Such an embedding method is different from the famous delay-embedding (Takens, 1981), which is an operation performed in signal's original space rather than dual space here. Delay-embedding could be vulnerable to short and noisy process. A messy point cloud could provide nothing except for 'topological noises'. However, by ranking power spectrum, data points are well-organized in an interpretable way and TDA could circumvent such pitfalls encountered in original space.

In conclusion, order structures of physiological signals' power spectra are almost neglected in existing methods but they are not meaningless. On the contrary, such structures could provide a unique perspective to understand the intrinsic properties of physiological processes.

## Acknowledgements

This work is supported by China Minister of Science and Technology and EU Commission collaborative project (No. 2017YFE0112000) and Shanghai Municipal Science and Technology Major Project (No. 2017SHZDZX01).

**Figures in the manuscript.**

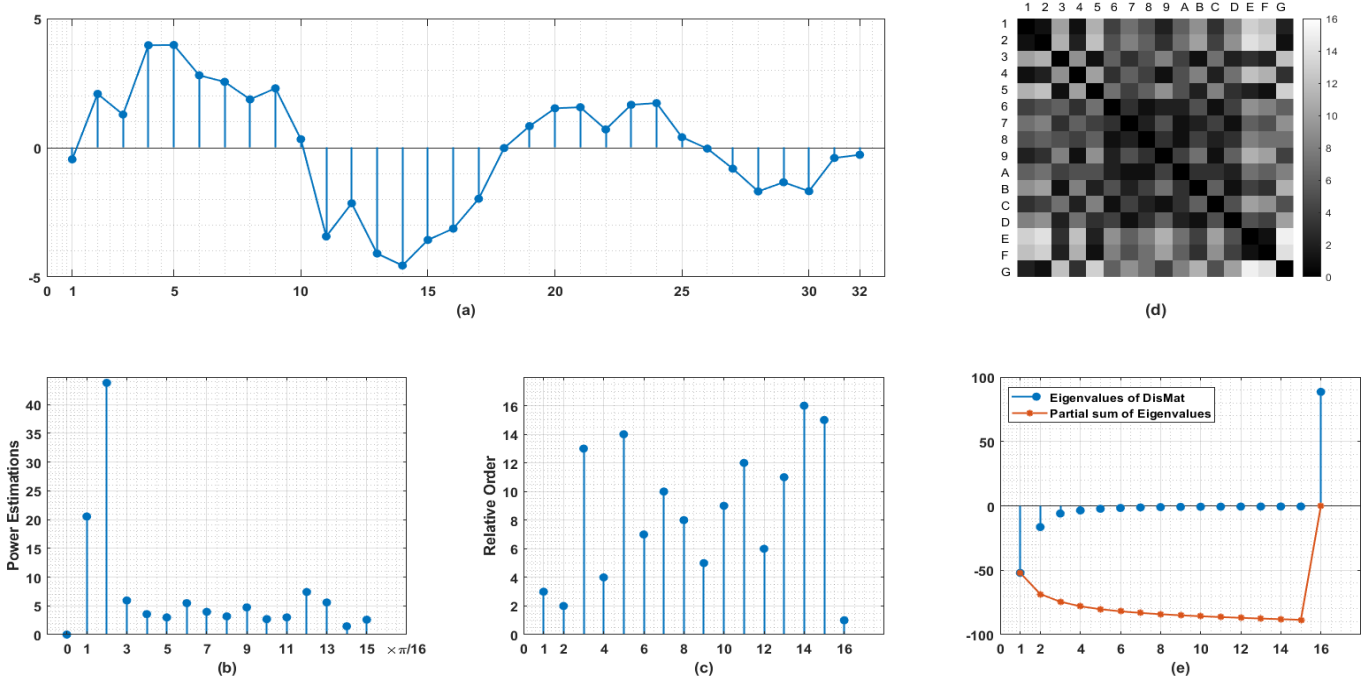

Figure.1. Schematic diagram of proposed method. (a) Original signal represented as equally-spaced time series. (b) Power estimations on normalized frequency range $[0, \pi]$, the number of points in (b) is just the half of that in (a). (c) Relative orders of frequency components. The values on horizontal axis indicate the ranking of frequency components in descending order. The values on verticle axis indicate the location of corresponding frequency component in normalized range $[0, \pi]$. (d) Distance matrix (DisMat) of frequency components arranged in (c) whereby A-G represent $10_{th}$-$16_{th}$ points. (e) Eigenvalues of DisMat and their partial sums. Note that DisMat's trace identically equals to 0.

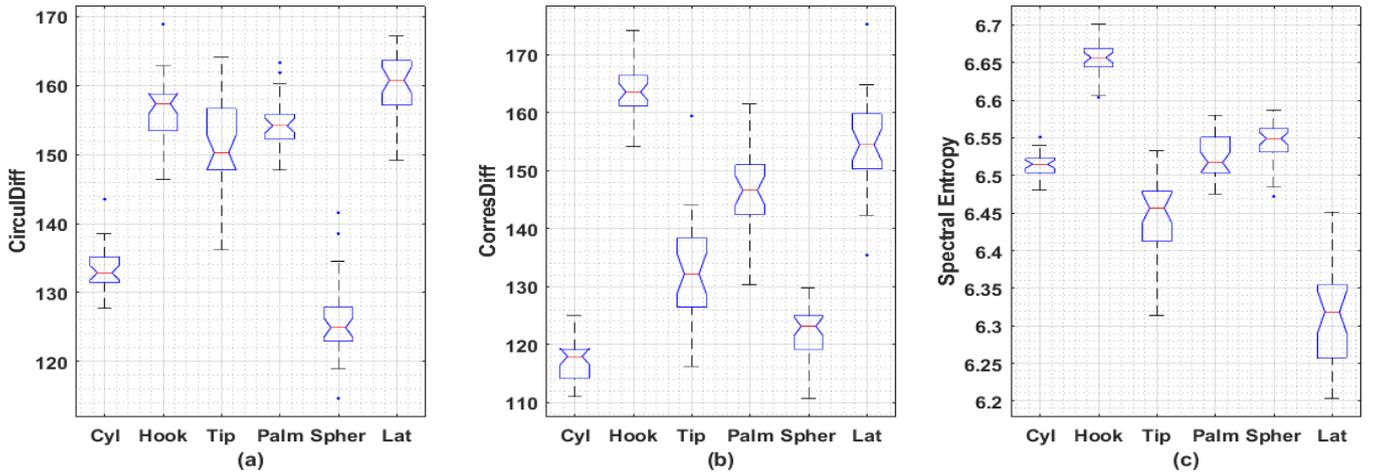

Figure. 2. Boxplots of irregularity metrics obtained from sEMG signals collected from 6 different actions. (a) Circular Difference Descriptor; (b) Correspondence Difference Descriptor; (c) Spectral Entropy; Totally there are 5 subjects (3 females and 2 males) involved in this dataset. Each subject repeated 30 times for each action with 2 channels' EMG signals recorded. Each recording is of length 6 seconds under sampling frequency 500 Hz (3000 data points). Only the latest 2048 data points for each recording are used to generate these metrics. Female 3, channel 2 is used here with remaining boxplots left in the Supplementary Information. Significant differences ($p<0.001$) between sample medians are observed among inter-subset Wilcoxon rank sum tests, except for Hook-vs-Tip ($p=0.0017$), Hook-vs-Palm ($p=0.0378$), Hook-vs-Tip ($p=0.0042$), Tip-vs-Palm ($p=0.0138$) in CirculDiff's comparisons and Cyl-vs-Palm ($p=0.2340$), Palm-vs-Spher ($p=0.0117$) in Spectral Entropy's comparisons. (Cyl: Cylindrical, for holding cylindrical tools; Hook: for supporting a heavy load; Tip: for holding small tools; Palm: Palmar, for grasping with palm facing the object; Spher: Spherical, for holding spherical tools; Lat: Lateral: for holding thin, flat objects; CirculDiff: Circular Difference Descriptor; CorresDiff: Correspondence Difference Descriptor).

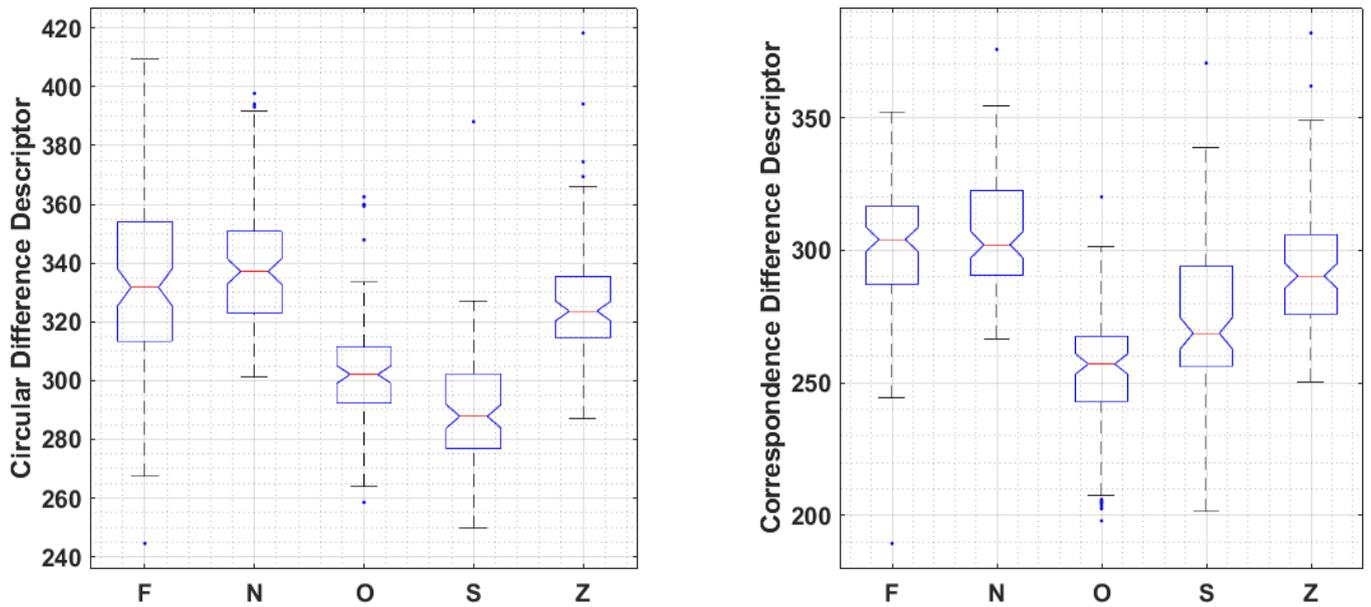

Figure.3. Boxplots of the value distributions of two descriptors. Bonn seizure dataset consists EEG recordings collected from 5 healthy volunteers with eyes open (Z) or closed (O), and 5 patients during ictal period (S), inter-ictal period in epileptogenic zone (F) or hippocampal formation of the opposite hemisphere (N). Another classification criterion is healthy EEG (OZ), inter-ictal period activities (FN) and ictal period activities (S). Significant differences ($p<0.001$) between sample medians are observed among inter-subset Wilcoxon rank sum tests, except for F-vs-N ($p=0.238$) and F-vs-Z ($p=0.002$) in Correspondence Difference Descriptor's comparisons, F-vs-N (for $p=0.158$) in Circular Difference Descriptor's comparisons.

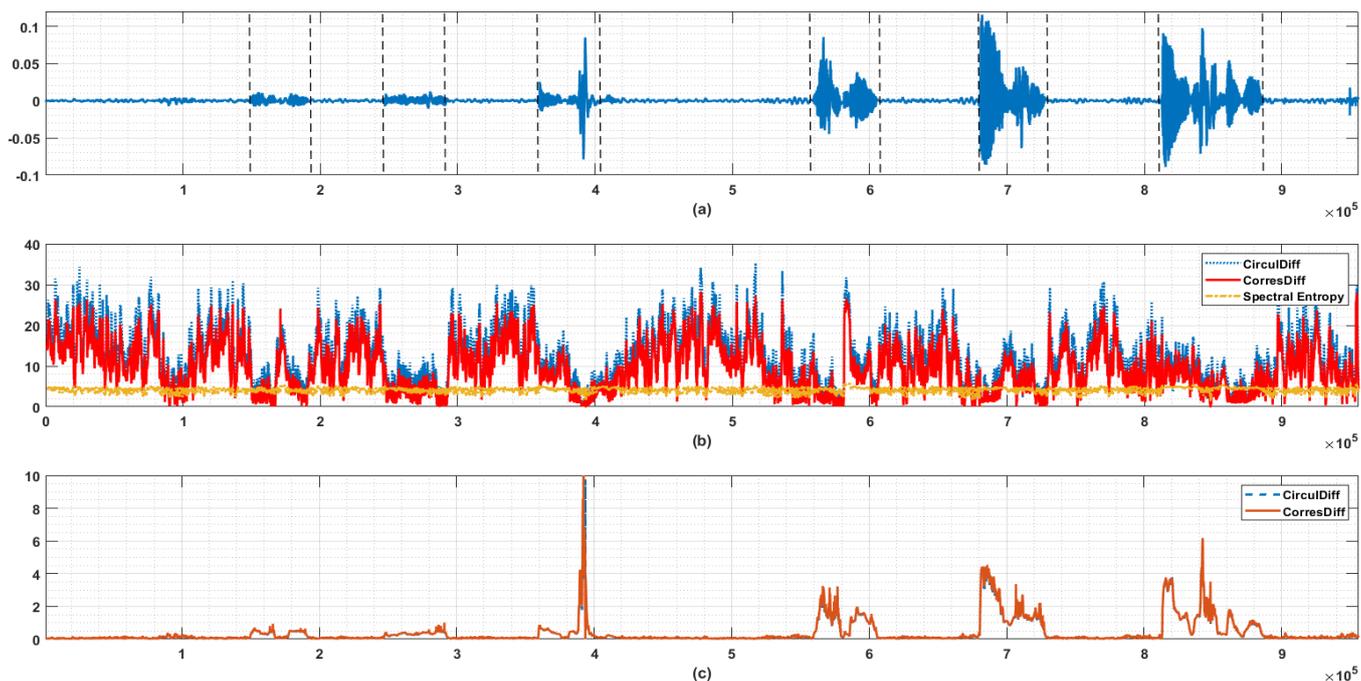

Figure. 4. Use case of endpoint detection in speech signal. (a) Original speech signal. The starts and end points of syllables are indicated by dashed lines. (b) Circular Difference Descriptor, Correspondence Difference Descriptor and Spectral Entropy of signal in (a). The window length is 1024 and step length is 128 here. The energy threshold $q$ is set to be 0.9 here. It is obvious that the proposed descriptors are more sensitive to the appearance of syllables (reduced irregularity implies possible formant caused) than spectral entropy. (c) Monitoring value taking local energy into consideration (let the standard deviation of last 1024 points be LE here). The values on y-axis are calculated by $\log_{10}(1 + LE)/\log_{10}(\text{Descriptor})$. (CirculDiff: Circular Difference Descriptor; CorresDiff: Correspondence Difference Descriptor).

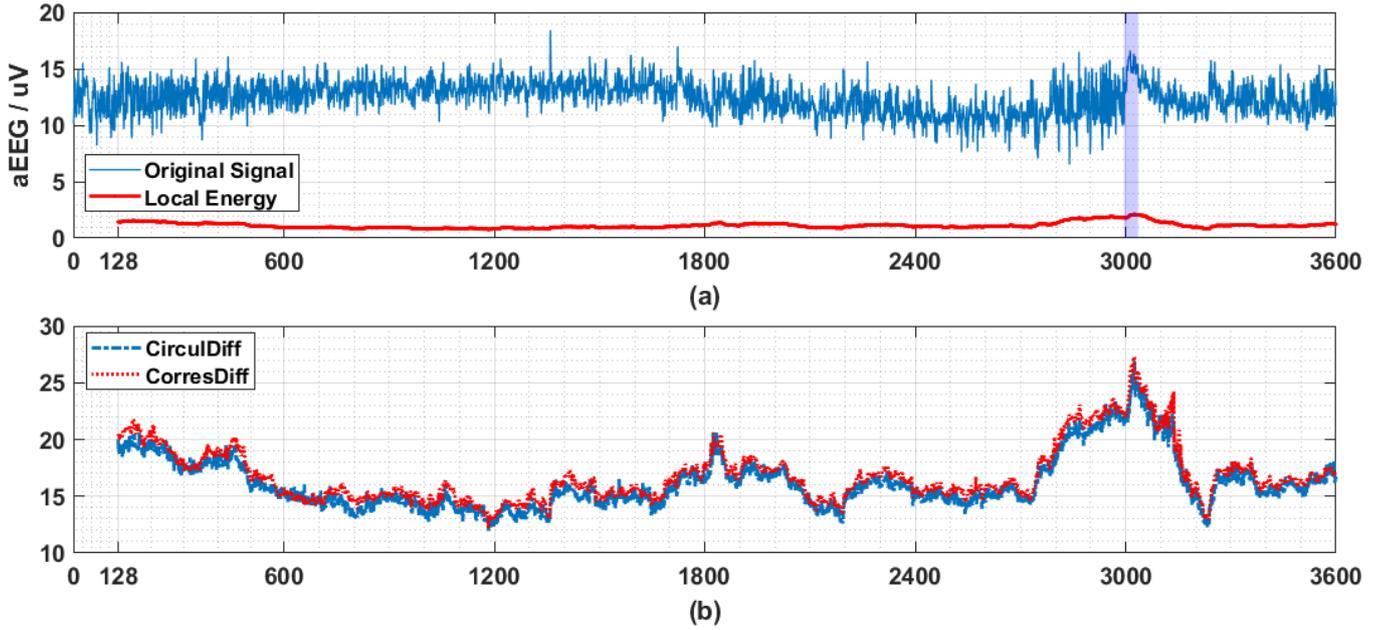

Figure. 5. The application of proposed descriptor on seizure detection. (a) Amplitude-integrated EEG (aEEG) tracing contains a seizure event (indicated by purple bar from $2996_{th}$ to $3036_{th}$ seconds) with its local energy (here standard deviation of last 128 points, referred as *LE* is used). The original multichannel EEG signals (23 channels with sampling frequency being 256 Hz) are drawn from patient 1, record 3 from CHB-MIT dataset and consist of one hour's recording. The channel 7 (C3-P3) is transformed by recommended pipeline into aEEG. Every second's data points (256 data points) are compreesed into 2 data points (upper bound and lower bound each). So totally 3600×2 data points are obtained. Here we use only upper bound and similar results have also been observed for lower bound data. (b) Circular Difference Descriptor and Correspondence Difference Descriptor monitoring (calculated based on last 128 data points in (a)). The step length is fixed to be 1 here, which means that after obtaining every new point, the values of two descriptors will be updated to provide the finest and most sensitive anomaly monitoring. The values on y-axis are calculated by $\log_{10}(1 + LE)/\log_{10}(Descriptor)$. Drastic change accompanied by the onset of seizure (a high value first and then a steep decresing) is observed in both of the descriptors (CirculDiff: Circular Difference Descriptor; CorresDiff: Correspondence Difference Descriptor).

**Supplementary Information (SI)**
1. Empirical distribution of two descriptors defined in (3) and (4)

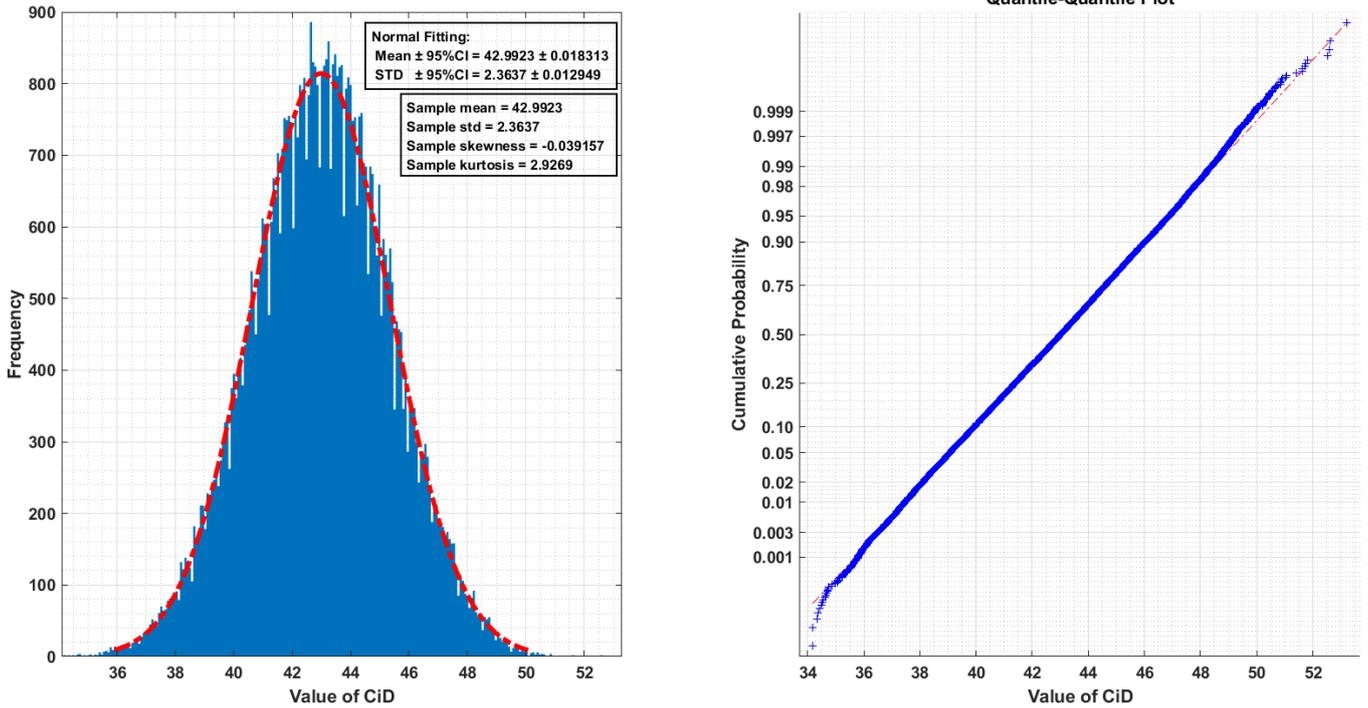

Figure. s1. Given $N = 64$, 640 000 times random generations of permutation over $N$ letters yields an empirical distribution of *Circular Difference Descriptor* ($C_iD$) fitting normal distribution well (red dot dash line) in the left panel (STD: standard deviation, CI: confidence interval). The right panel is the quantile-quantile plot of this empirical distribution (blue cross) with theoretical value of normal distribution (red dot dash line).

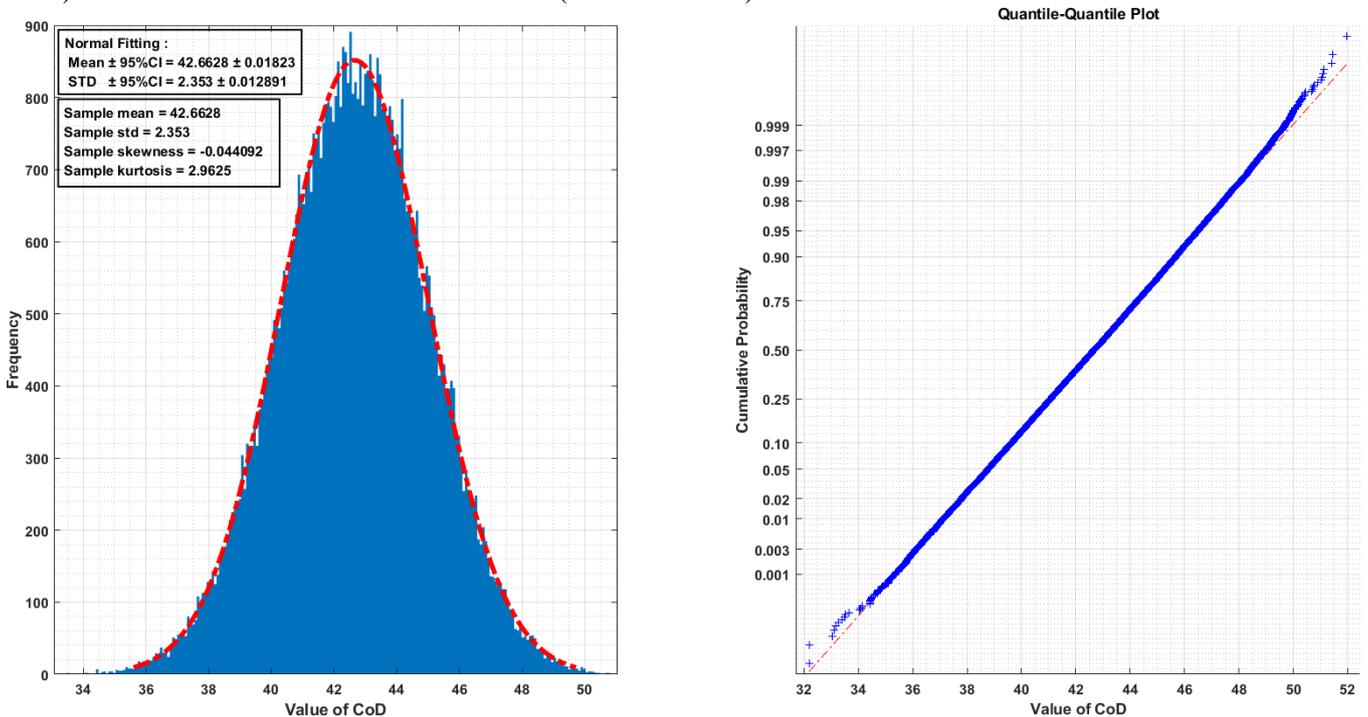

Figure. s2. Given $N = 64$, 640 000 times random generations of permutation over $N$ letters yields an empirical distribution of *Correspondense Difference Descriptor* ($C_oD$) fitting normal distribution well (red dot dash line) in the left panel (STD: standard deviation, CI: confidence interval). The right panel is the quantile-quantile plot of this empirical distribution (blue cross) with theoretical value of normal distribution (red dot dash line).

2. Full intra-subject comparison of two descriptors across six different actions.

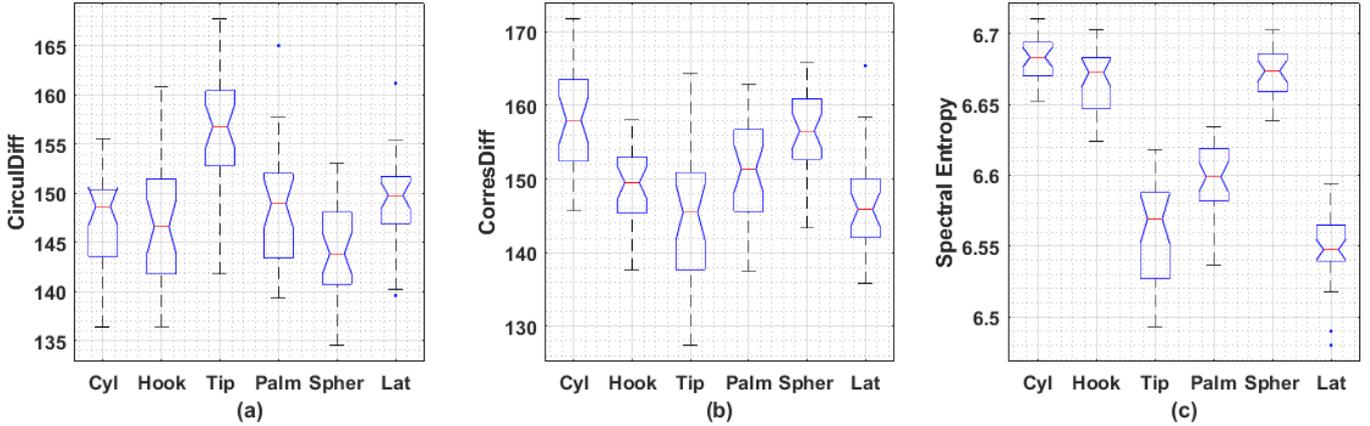

Figure. s3. Comparisons of female one, channel one. Cases failed to pass Wilcoxon rank sum tests (13/45, confidence level is set to be 0.05): Cyl-vs-Hook (p=0.4204), Cyl-vs-Palm (p=0.5997), Cyl-vs-Lat (p=0.1297), Hook-vs-Palm (p=0.2519), Hook-vs-Spher (p=0.1882), Palm-vs-Lat (p=0.4733) in (a); Cyl-vs-Spher (p=0.5395), Hook-vs-Tip (p=0.0555), Hook-vs-Palm (p=0.3255), Hook-vs-Lat (p=0.0850), Tip-vs-Lat (p=0.4918) in (b); Hook-vs-Spher (p=0.3329), Tip-vs-Lat (p=0.3042) in (c).

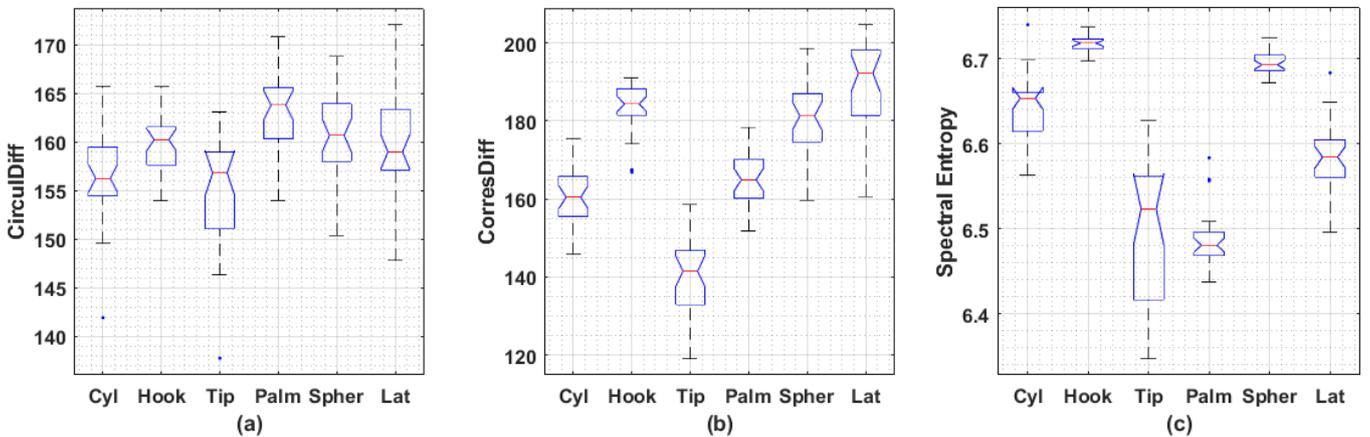

Figure. s4. Comparisons of female one, channel two. Cases failed to pass Wilcoxon rank sum tests (6/45, confidence level is set to be 0.05): Cyl-vs-Tip (p=0.4247), Hook-vs-Spher (p=0.2116), Hook-vs-Lat (p=0.9117), Spher-vs-Lat (p=0.4204) in (a); Hook-vs-Spher (p=0.1413) in (b); Tip-vs-Palm (p=0.2707) in (c).

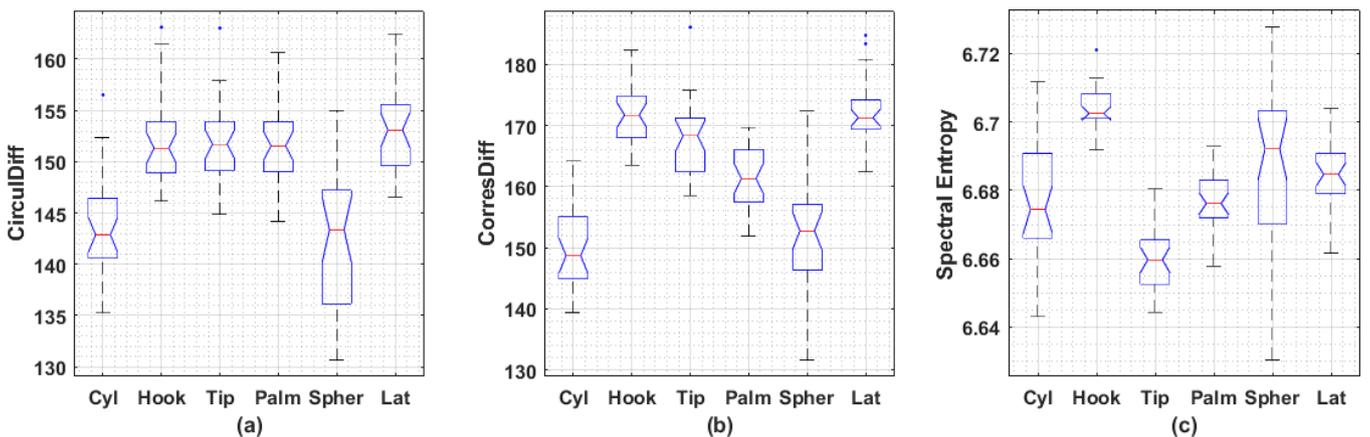

Figure. s5. Comparisons of female two, channel one. Cases failed to pass Wilcoxon rank sum tests (12/45, confidence level is set to be 0.05): Cyl-vs-Spher (p=0.8303), Hook-vs-Tip (p=0.8418), Hook-vs-Palm (p=0.9000), Hook-vs-Lat (p=0.3711), Tip-vs-Palm (p=0.9352), Tip-vs-Lat (p=0.2340), Palm-vs-Lat (p=0.3042) in (a); Cyl-vs-Spher (p=0.2838), Hook-vs-Lat in (b); Cyl-vs-Palm (p=0.9941), Cyl-vs-Spher (p=0.1224), Cyl-vs-Lat (p=0.0850), Spher-vs-Lat (p=0.4204) in (c).

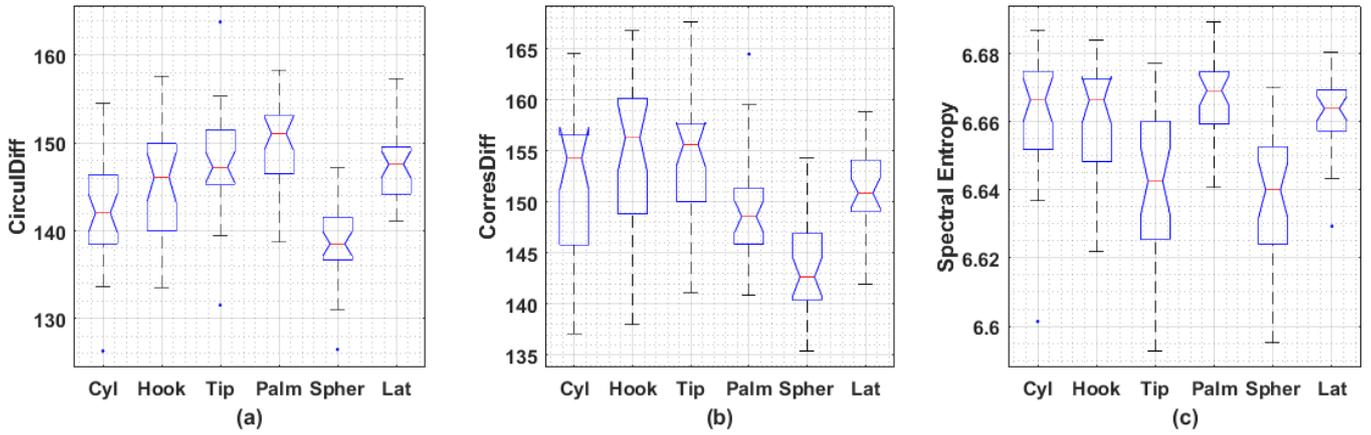

Figure. s6. Comparisons of female two, channel two. Cases failed to pass Wilcoxon rank sum tests (17/45, confidence level is set to be 0.05): Hook-vs-Tip (p=0.1958), Hook-vs-Lat (p=0.2838), Tip-vs-Palm (p=0.0555), Tip-vs-Lat (p=0.5298) in (a); Cyl-vs-Hook (p=0.2009), Cyl-vs-Tip (p=0.3555), Cyl-vs-Palm (p=0.0657), Cyl-vs-Lat (p=0.1494), Hook-vs-Tip (p=0.5011), Palm-vs-Lat (p=0.0690) in (b); Cyl-vs-Hook (p=0.6414), Cyl-vs-Palm (p=0.4204), Cyl-vs-Lat (p=0.6414), Hook-vs-Palm (p=0.2340), Hook-vs-Lat (p=0.9941), Tip-vs-Spher (p=0.4035), Palm-vs-Lat (p=0.1260) in (c).

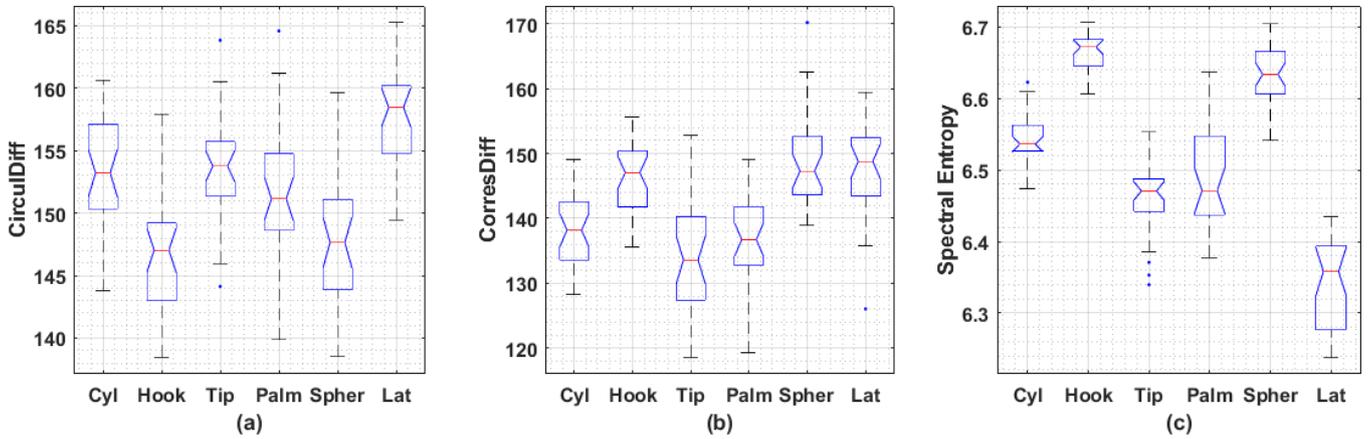

Figure. s7. Comparisons of female three, channel one. Cases failed to pass Wilcoxon rank sum tests (10/45, confidence level is set to be 0.05): Cyl-vs-Tip (p=0.6735), Cyl-vs-Palm (p=0.2062), Hook-vs-Spher (p=0.3112), Tip-vs-Palm (p=0.0575) in (a); Cyl-vs-Palm (p=0.4204), Hook-vs-Spher (p=0.4643), Hook-vs-Lat (p=0.5201), Tip-vs-Palm (p=0.2707), Spher-vs-Lat (p=0.7506) in (b); Tip-vs-Palm (p=0.3790) in (c).

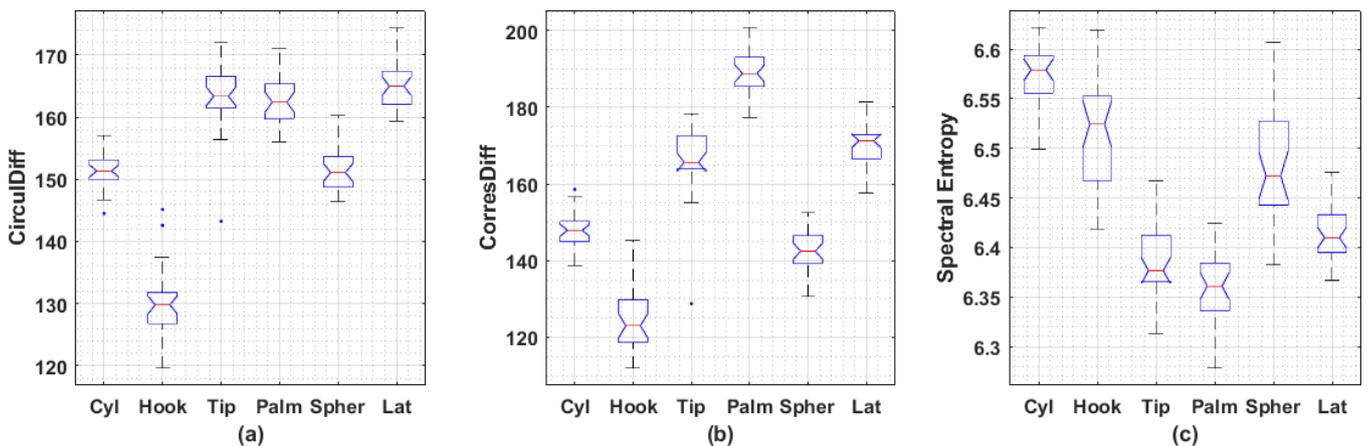

Figure. s8. Comparisons of male one, channel one. Cases failed to pass Wilcoxon rank sum tests (4/45, confidence level is set to be 0.05): Cyl-vs-Spher (p=0.7394), Tip-vs-Palm (p=0.2282), Tip-vs-Lat (p=0.5011) in (a); Tip-vs-Lat (p=0.1297) in (b).

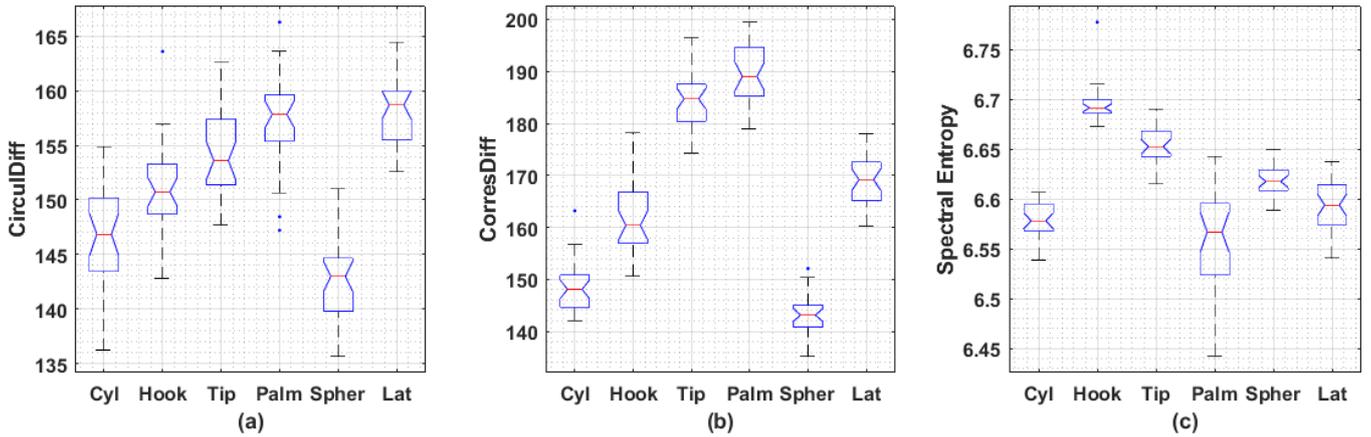

Figure. s9. Comparisons of male one, channel two. Cases failed to pass Wilcoxon rank sum tests (2/45, confidence level is set to be 0.05): Palm-vs-Lat (p=0.2838) in (a); Cyl-vs-Palm (p=0.2643) in (c).

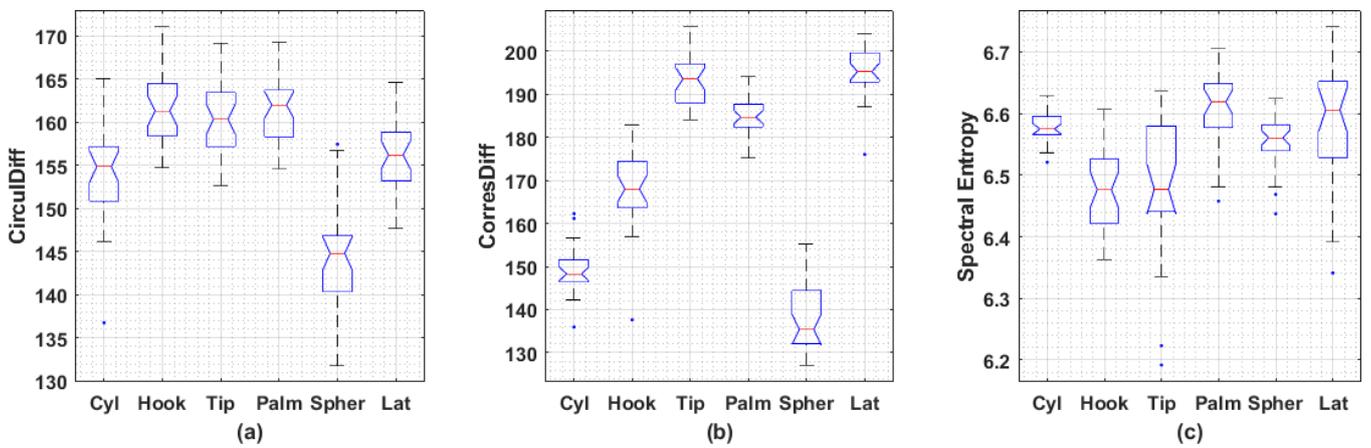

Figure. s10. Comparisons of male two, channel one. Cases failed to pass Wilcoxon rank sum tests (8/45, confidence level is set to be 0.05): Cyl-vs-Lat (p=0.0993), Hook-vs-Tip (p=0.3329), Hook-vs-Palm (p=0.9705), Tip-vs-Palm (p=0.4553) in (a); Tip-vs-Lat (p=0.0724) in (b), Cyl-vs-Lat (p=0.1715), Hook-vs-Tip (p=0.5997), Palm-vs-Lat (p=0.5895) in (c).

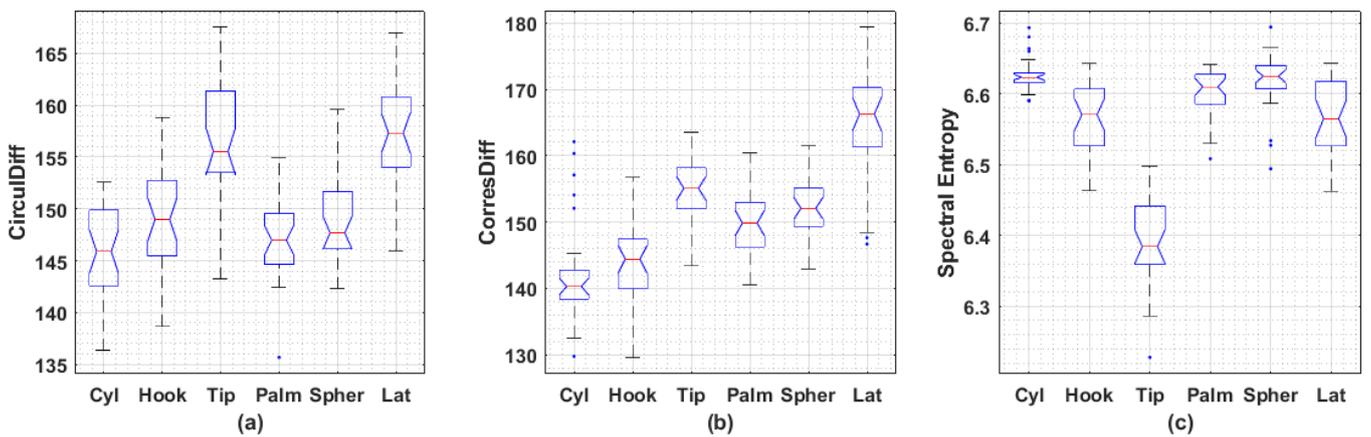

Figure. s11. Comparisons of male two, channel two. Cases failed to pass Wilcoxon rank sum tests (8/45, confidence level is set to be 0.05): Cyl-vs-Palm (p=0.3953), Hook-vs-Palm (p=0.0933), Hook-vs-Spher (p=0.6520), Tip-vs-Lat (p=0.6843), Palm-vs-Spher (p=0.1669) in (a); Cyl-vs-Hook (p=0.0724) in (b), Cyl-vs-Spher (p=0.9000), Hook-vs-Lat (p=0.9941) in (c).